\documentstyle[12pt,psfig]{article}
\begin{document}

\title {\textsf{The Spherical Relativistic Detonation of Scalaron Stars}}
\author { V.~Folomeev,
V.~Gurovich \thanks{email: astra@freenet.kg} \enspace and
R.~Usupov
\\
{\small \it  Physics Institute of NAN KR,}\\
         {\small \it 265 a, Chui str., Bishkek, 720071,  Kyrgyz Republic.}}
\date{\small \it (6 September 2000)}
\maketitle

\begin{abstract}
\noindent Now the hypothesis of existence of scalar fields of a
various nature and energy density in the modern Universe is
intensively explored. It can explain a nature of the dark (non-
baryon) matter in the Universe and an existence of positive
$\Lambda $-term~\cite{ref:Sahni}. One of component of such field
has a cluster nature and organizes in the closed gravitational
configurations from galactic scales up to relativistic microscopic
stars. In the authors paper~\cite{ref:Fol}  the hypothesis of
detonation of such fields was considered. As a result of phase
transition behind the wavefront a relativistic plasma of high
energy density can appear. This process is similar to a
relativistic detonation and it can create macroscopic fireballs
sufficient for an explanation of the phenomenon of gamma-ray
bursts~\cite{ref:Piran}. In Ref.~\cite{ref:Fol} it was supposed
that the front of such ''detonation'' wave is entered by the flow
of scalar fields with constant energy density. If the size of the
formed plasma configuration is commensurable with the size of
scalaron cluster, this hypothesis is not correct. It is necessary
to take into account a modification of the energy density of the
scalar field from centre to a periphery. It is changes the
dynamics of the fireball on principle. The indicated problem in
framework of special relativity is considered in this paper.
\end{abstract}

\section{Introduction}

\noindent Until recently the problems of a relativistic detonation
when a velocities both of wave and gas (plasma) achieve of
near-light velocities had only methodical interest. In connection
with the hypothesis of a detonation regime of ''burning'' of the
cosmic scalar fields~\cite{ref:Fol} this problem takes on physical
interest. As well as in Ref.~\cite{ref:Fol} we suppose that there
is a scalaron star in which the pressure of the field is
equilibrated by the weak gravitational forces (Newtonian
approximation). Then it is possible to describe the specified
process of detonation in framework of special relativity. Other
field of application can be, in principle, the processes of
laboratory scale~\cite{ref:Lasers}.

At description of this phenomenon at the front of ''detonation''
wave according to conservation laws both of relativistic momenta
density and momenta density flow  $T_{0}^{1}$(field) = $T_{0}^{1}$(plasma)
and $T_{1}^{1}$(field) = $%
T_{1}^{1}$(plasma) for the scalar field in the rest, on which the
wavefront with velocity $\mathcal{D}$\ is passing, we have:
\begin{equation}
\label{eq1}
{\mathcal{D}}=\frac{2v_{p}}{1+v_{p}^{2}};\,\,\,\,\,\,\,\,\,\,\,\varepsilon
_{p}=\frac{2}{1-\omega ^{2}}\varepsilon _{f}\,\,.
\end{equation}
Here, $\varepsilon _{f}$ is the energy density of the scalar field
in scalaron regime (it forms from density both of potential and
kinetic energy of the field), $\varepsilon _{p}$\ is the energy
density of plasma behind the wavefront. Thus the plasma removes
from the wavefront with the velocity of sound $\omega $\ that
corresponds to a condition of normal detonation. As well as in
Ref.~\cite{ref:Fol} we choose the scalar field of the simplest
form
\begin{equation}
\label{eq2}
T_{i}^{k}=\varphi _{,i}\varphi ^{,k}-\delta _{i}^{k}\left( \frac{1}{2}%
\varphi _{,\mu }\varphi ^{,\mu }-V(\varphi )\right)
,\,\,\,\,\,\,\,\,V(\varphi )=m^{2}\varphi ^{2}/2\,,
\end{equation}
which has rather small spatial derivatives. In Eqn. (\ref{eq1})
$v_{p}$\ is a velocity of high-temperature plasma relative to the
wavefront $\mathcal{D}$.

\section{The Homogeneous Detonation}

In this report we are interesting in dynamics of the detonation
wave and the current behind one in the case when $\varepsilon
_{f}$\ depends on radial coordinate $r$. For considering of this
problem it is expediently to repeat briefly the results of the
homogeneous case. The set of equations of relativistic
gas-dynamics in special relativity for spherical frame with use of
usual radial velocity $v$\ is possible to write as:
\begin{eqnarray}
\label{eq3}
\frac{1}{\theta ^{2}}\left( \frac{\partial v}{\partial \tau }+v\frac{%
\partial v}{\partial r}\right) +\frac{1}{w}\left( \frac{\partial p}{\partial
r}+v\frac{\partial p}{\partial \tau }\right)  &=&0,  \nonumber \\
\frac{1}{w}\left[ \frac{\partial \varepsilon }{\partial \tau }+v\frac{%
\partial \varepsilon }{\partial r}\right] +\frac{1}{\theta ^{2}}\left( \frac{%
\partial v}{\partial r}+v\frac{\partial v}{\partial \tau }\right) +\frac{2v}{%
r} &=&0, \\
\frac{\partial \sigma }{\partial \tau }+v\frac{\partial \sigma }{\partial r}%
+\sigma \left[ \frac{1}{\theta ^{2}}\left( \frac{\partial v}{\partial r}+v%
\frac{\partial v}{\partial \tau }\right) +\frac{2v}{r}\right]  &=&0.
\nonumber
\end{eqnarray}
Here, $\sigma $ is the entropy density, $\theta ^{2}=1-v^{2}$,
$w=\varepsilon +p$ and $c=1$ . According to the classical theory
of a spherical detonation considered by
Zel'dovich~\cite{ref:Land}, the pattern of current is self-similar
and depending from the unique variable
\begin{equation}
\label{eq4}
\xi =r/\tau .
\end{equation}
For the relativistic problem we maintain the same self-similarity.
The mentioned above velocity $v_{p}$\ relative to the wavefront
should be equal to the velocity of sound of relativistic plasma
$\omega =1/\sqrt{3}$. Thus the velocity of the detonation wave
${\mathcal{D}}=\sqrt{3}/2$\ according to Eqn. (\ref{eq1}). The set
of Eqns. (\ref{eq3}) supposes the deriving of one equation
relative to velocity $v$:
\begin{equation}
\label{eq5}
\frac{dv}{d\xi }\left[ \frac{1}{\omega ^{2}}\left( \frac{v-\xi }{1-v\xi }%
\right) ^{2}-1\right] =\frac{2v}{\xi }\frac{\theta ^{2}}{1-v\xi }.
\end{equation}
The analysis of this equation is carried out similarly to
Ref.~\cite{ref:Land} and the diagrams of solutions both for $v$\
and energy density of plasma $\varepsilon _{p}$\ are shown in Fig.
1.
\begin{figure}
\begin{picture}(500,270)
\put(-30,-140){\includegraphics{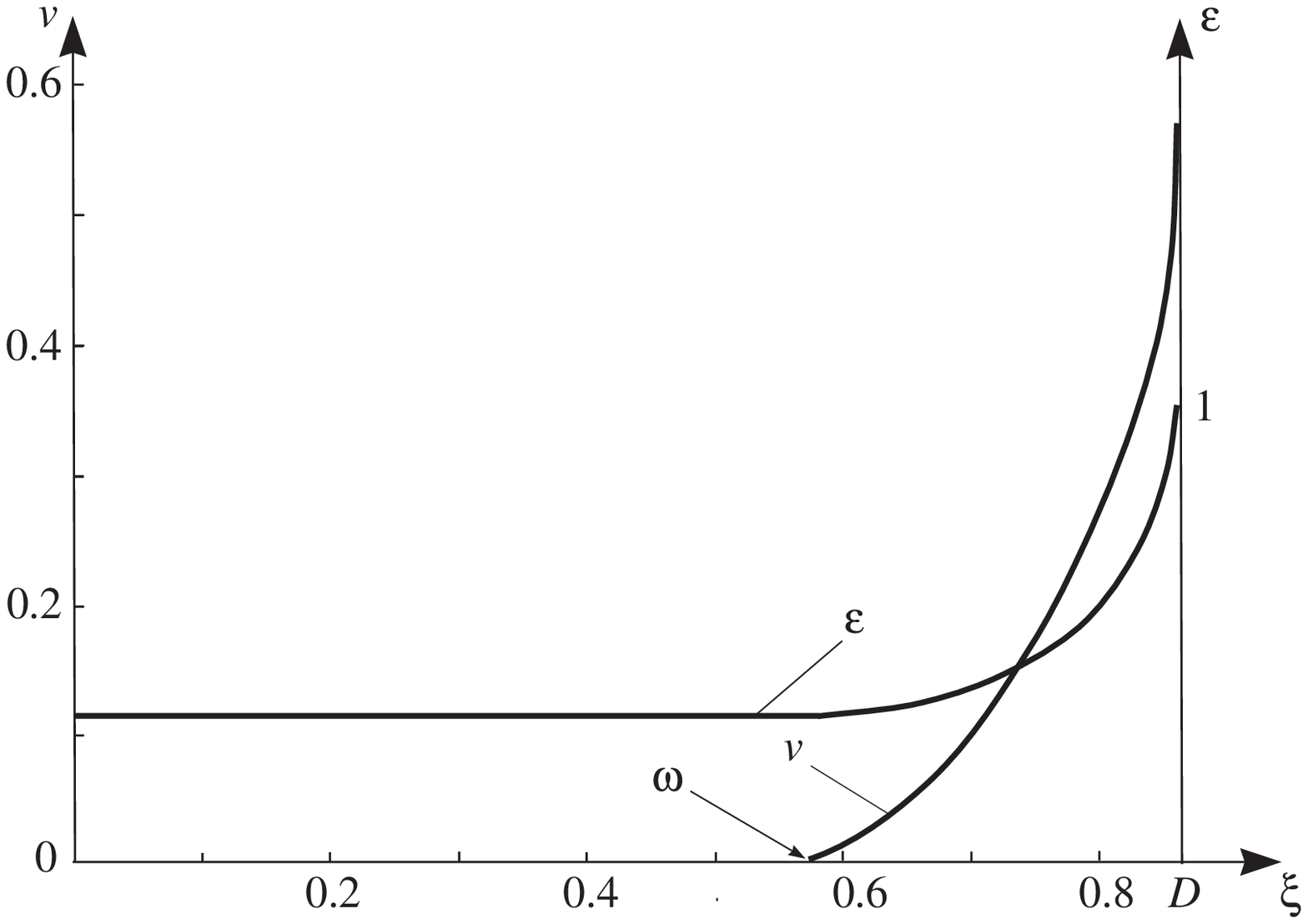}}
\end{picture}
\caption{\it The dependence of the energy density $\varepsilon$
and velocity $v$ on the self-similar variable $\xi$.}
\end{figure}
At the wavefront the derivative of the velocity tends to infinity.
At $\xi =\omega $\ the solution $v(\xi )$\ is matched with $v=0$\
with maintenance of a continuity of first derivative. This point
corresponds to a contact discontinuity. Thus the solution for
$\varepsilon _{p}(\xi )$\ is obtained from the second
equation of system (\ref{eq3}). It is easy to show that for photon gas with $%
p=\varepsilon /3$\ and entropy density $\sigma \sim \varepsilon
_{p}^{3/4}$~\cite{ref:Land} the third equation from (\ref{eq3}) is
satisfy identically. It is necessary to note that the
nonrelativistic analog of this problem, considered for the first
time by Zel'dovich, corresponds to isentropic flow of classical
gas-dynamics. As it was specified above, it is not valid for a
relativistic problem where the flow is adiabatic. In
paper~\cite{ref:Fol} the statement about isentropy in the
relativistic problem is incorrect, though all obtained results are
valid by virtue of the mentioned above note about identical
realization of the equation for the entropy from (\ref{eq3}).

\section{The Inhomogeneous Detonation}

\noindent As the scalar field $\varepsilon _{f}$ which is
''burnt'' in the detonation wave represents a gravitational
cluster (scalaron ''star''), its energy depends on coordinate $r$.
In the example of solution considered in Ref.~\cite{ref:Fol} for
such cluster $\varepsilon _{f}\sim r^{-2}$\ \ except for central
area. In this connection it is interesting to find a self-similar
solution for the relativistic detonation at $\varepsilon _{f}(r)$.
Note that the velocity of the detonation wave according to Eqn.
(\ref{eq1}) does not depend on energies density $\varepsilon
_{f},\varepsilon _{p}$\ and remains constant. On this reason the
mentioned above dependence is possible to present as:
\begin{equation}
\label{eq6}
\varepsilon _{f}=\left( \frac{\tau _{0}}{\tau }\right)
^{n}E(\xi ),
\end{equation}
where $\xi $\ is still determined by expression (\ref{eq4}) and
$E(\xi )$\ is a representative of the function $\varepsilon _{f}$.
In this case also is possible to obtain one equation on $v(r)$\
which is possible to write as
\begin{equation}
\label{eq7}
\frac{dv}{d\xi }\left[ \frac{1}{\omega ^{2}}\left( \frac{v-\xi }{1-v\xi }%
\right) ^{2}-1\right] =\frac{2v}{\xi }\frac{\theta ^{2}}{1-v\xi }-\frac{%
n\theta ^{4}}{(1+\omega ^{2})(1-v\xi )^{2}}.
\end{equation}
{\looseness=-2 (The equation for the entropy is still satisfy identically.) At
$n=0$\ the last equation passes in Eqn. (\ref{eq5}). The key
difference of this equation from Eqn. (\ref{eq5}) is the absence
of solution $v=0$\ that changes qualitatively the form of solution
shown in Fig. 1. On the other hand the small values of $n$\ can
not in essence change the form of the solution near to front. It
is necessary to expect that there is a critical value $n_{\ast }$\
since which this region of solution will be essential change also.
It can be found from the following reasons: infinite derivative
$dv/d\xi $\ at the wavefront, as well as in Eqn. (\ref{eq5}) (see
Fig. 1), is achieved at tendency of the factor in brackets on the
{\it l.h.s.} in Eqn. (\ref{eq7}) to zero from above and positive
determinancy of expression on the {\it r.h.s.} At realization of
these requirements the flow behind the front will coincide
qualitatively with the case considered already~\cite{ref:Fol}. The
differences are appeared that
on the side of small values of $v$\ in central region the solution $v(\xi )$%
will grow linearly from centre and the complete solution is
obtained by matching of the specified two solutions. Thus the
value of velocity remains continuous and the jump is experienced
only by derivatives (Fig. 2).
\begin{figure}
\begin{picture}(500,270)
\put(-20,-390){\includegraphics{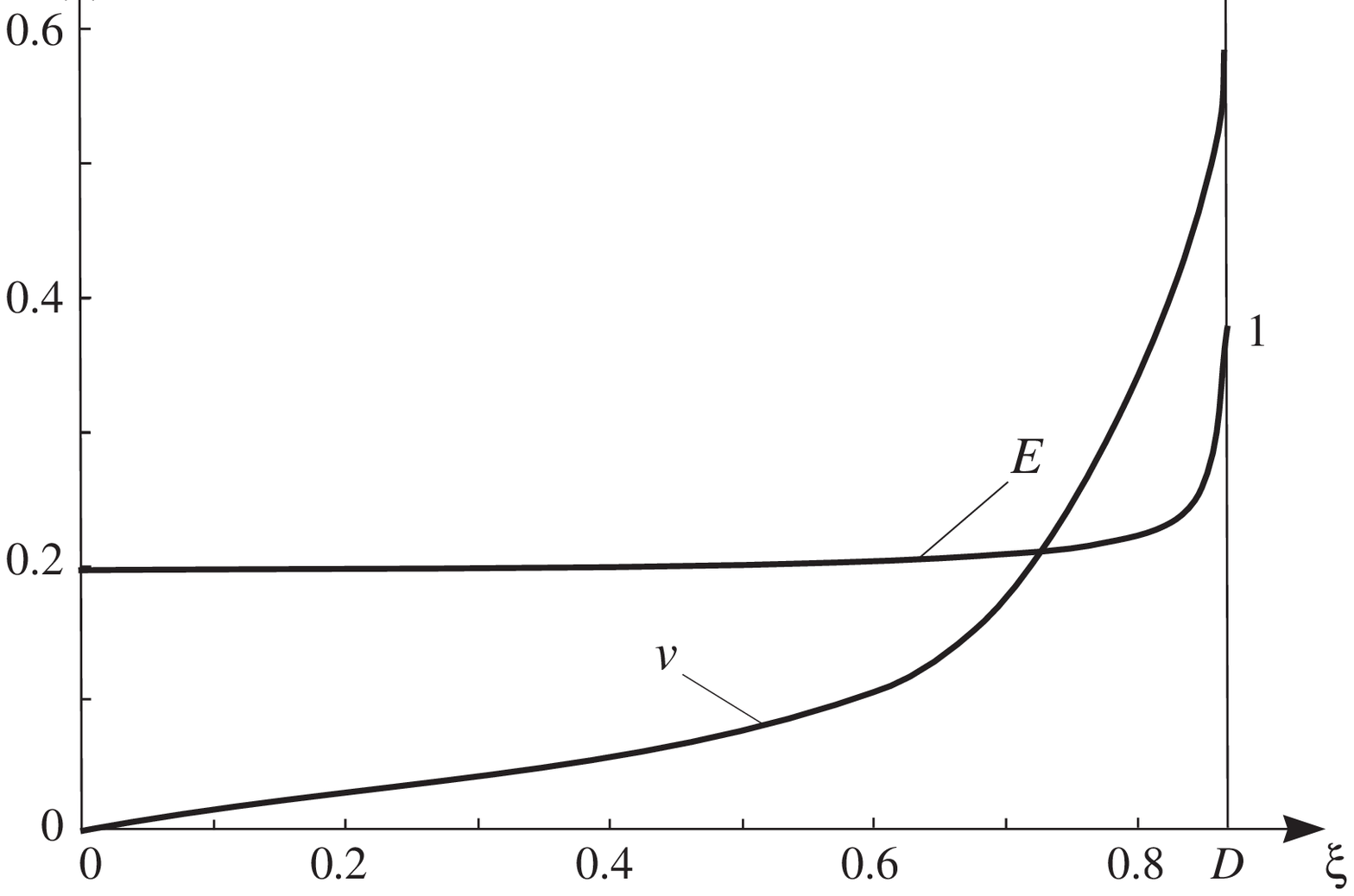}}
\end{picture}
\caption[Fig1]{\it The dependence of the representative of energy
density $E$ and velocity $v$ on the self-similar variable $\xi$ in
the case of n=0.5. The singular point (node) of Eqn. (\ref{eq7})
is in the point $\xi=0.66$.}
\end{figure}
Note that the point of matching is not casual. According to the
qualitative theory of the differential equations it corresponds to
a singular point - node. The branches of solutions $v(\xi )$\
going out from the point $v,\xi =0 $\ and from the wavefront at
$\xi =\mathcal{D}$\ and $v=\omega $\ with inevitability fall into
this point. Note that the centre ($\xi=0$) is a singular point - saddle
and the solution $v(\xi)$ in a neighbourhood
of centre which is interesting for us is a separatrix going from a saddle
to a node that is according
to the qualitative theory of the differential equations. In accordance
with increasing of $n$\ the coordinate
of this node on the axis $\xi $\ is monotonically moved from
the minimum value $n=0,\,\xi =\omega $\ up to $\xi =\mathcal{D}$\ at $%
n_{\ast }=1+\omega ^{2}=4/3$. It is a limiting value of
inhomogeneity of the energy density of the scalar field at which
the solution $v(\xi )$\ remains continuous from centre up to the
wavefront. Since $n>n_{\ast }$\ the condition of
''over-compressed'' detonation is realized~\cite{ref:Land} when
the velocity of the detonation wave becomes larger then the
specified value $\mathcal{D}$\ from Eqn. (\ref{eq1}). Behind the
wavefront the supersonic current takes place which terminates in a
relativistic shock wave
(SW). Behind the SW the solution is matched with continuous current of $v(\xi )$%
\ going up to centre. The coordinate of the SW and the parameters
of plasma are determined from conservation laws on the
SW~\cite{ref:Land}. This variant is interesting because of the
velocities both of the front $\mathcal{D}$\ and plasma behind it
can achieve ultra relativistic values (${\mathcal{D}},v\rightarrow
1$).

}

The case when the exponent $n<0$ has the special interest. Last
means that the energy density of the scalar field grows from
centre to periphery. (Note that for field stars, when the
effective negative pressure is realized, such variants are
possible.) The analysis of the Eqn. (\ref{eq7}) shows that as
against considered before cases the velocity of the plasma in a
neighbourhood of centre becomes negative (the motion to centre).
It easy to understand qualitatively, as the increasing density of
the field energy enters the wavefront. The adiabatic compression
of gas in central region brings to increase of the energy density
at centre (Fig. 3).
\begin{figure}
\begin{picture}(500,300)
\put(-30,-350){\includegraphics{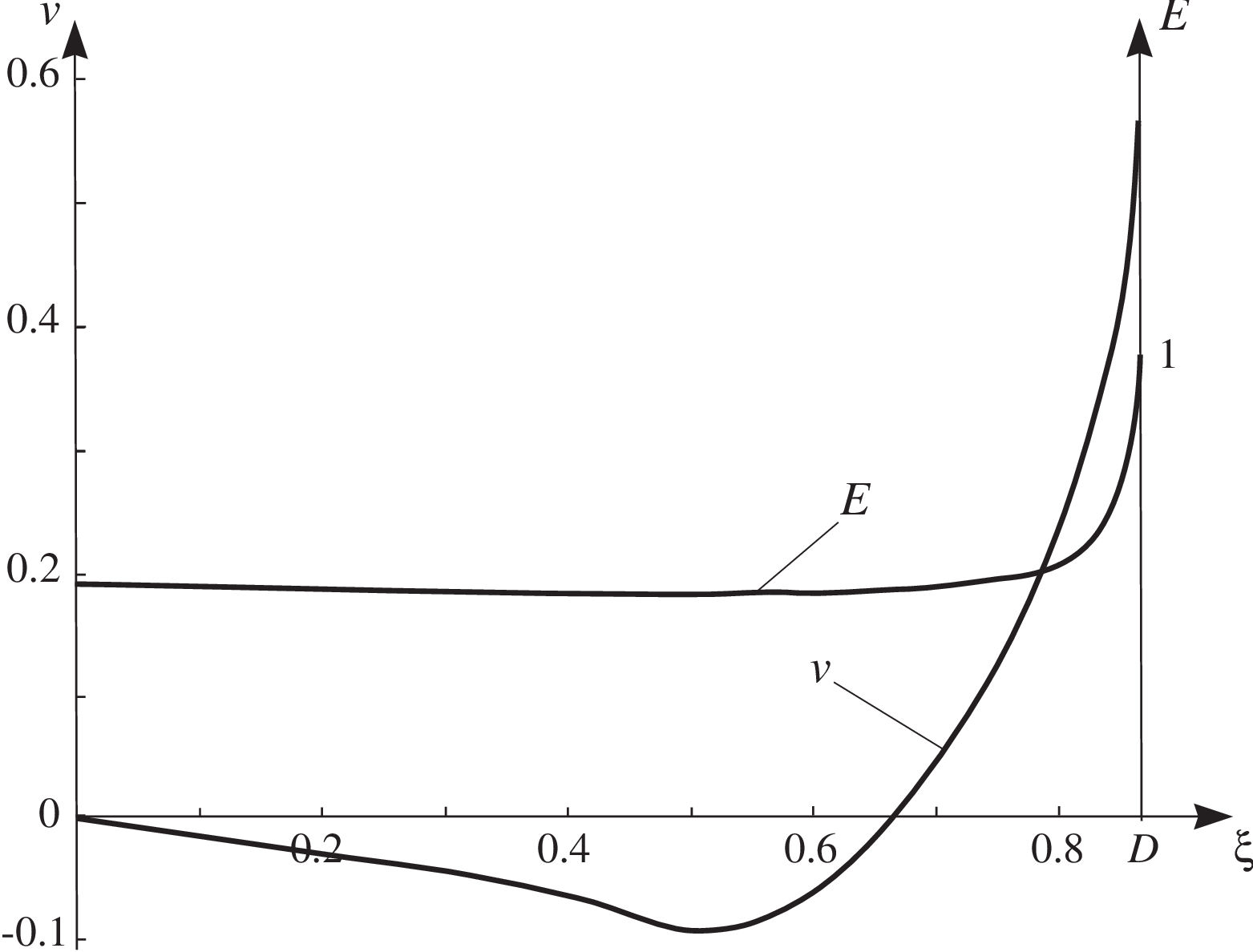}}
\end{picture}
\caption{\it The dependence of the representative of energy
density $E$ and velocity $v$ on the self-similar variable $\xi$ in
the case of n=-0.5. The singular point (node) of Eqn. (\ref{eq7})
is in the point $\xi=0.51$.}
\end{figure}

It is important to note that for laboratory problems the similar
control of the energy density by laser radiation can brings to
increase of temperature at centre. It can be important for
processes of laser detonation in a spherical
regime~\cite{ref:Lasers}.

For relativistic scalar stars the considered processes should be
described, certainly, in framework of general relativity.

\end{document}